# Non-Abelian Topological Phases and Their Quotient Relations in Acoustic Systems

Xiao-Chen Sun[1,2,§], Jia-Bao Wang[1,§], Cheng He[1,2,3,*], Yan-Feng Chen,[1,2,3,†]

[1]National Laboratory of Solid State Microstructures & Department of Materials Science and Engineering, Nanjing University, Nanjing 210093, China

[2]Collaborative Innovation Center of Advanced Microstructures, Nanjing University, Nanjing 210093, China

[3]Jiangsu Key Laboratory of Artificial Functional Materials, Nanjing University, Nanjing 210093, China

*Corresponding author.

chenghe@nju.edu.cn

†Corresponding author.

yfchen@nju.edu.cn

§ X.-C. S., J.-B. W. contributed equally to this work.

**Abstract**

Non-Abelian topological phases (NATPs) are highly sought-after candidate states for quantum computing and communication while lacking straightforward configuration and manipulation, especially for classical waves. In this work, we exploit novel braid-type couplings among a pair of triple-component acoustic dipoles, which act as functional elements with effective imaginary couplings. Sequencing them in one dimension allows us to generate acoustic NATPs in a compact yet reciprocal Hermitian system. We further provide the whole phase diagram that encompasses all *i*, *j*, and *k* non-Abelian phases, and directly demonstrate their unique quotient relations via different endpoint states. Our NATPs based on real-space braiding may inspire the exploration of acoustic devices with non-commutative characters.

Topological phases represent a novel form of matter [1–5] that has generated significant interest over the last decade for designing band topology in artificial materials, such as photonic and phononic crystals [6–16]. Their global behavior introduces robustness into energy bands in $k$-space, resulting in stable wave transmission against unfavorable but unavoidable perturbations or defects in practical fabrications. The construction or breaking of nodal points or lines of two neighboring bands typically plays an essential role. One well-known case is topological semimetals with topological charges, e.g., the $k$-sphere or $k$-loop in the Brillouin zone (BZ) enclosing a nodal point, which can manifest nontrivial Fermi arcs [17,18]. Breaking such local band degeneracy further can give rise to topological insulators, whose topological invariants such as Chern number [1], $Z_2$ invariant [19], and Zak phase [20] characterize the global property of isolated bulk bands. Then, nontrivial boundary states appear in the band gap. So far, they have usually been used to deal with one gap or nodal points between two adjacent bands, most of which belong to the Abelian group with commutative property, making them accumulated linearly.

Things could be very different with more than two bands that can interact with each other. For instance, in a three-band system, there are two interactive band gaps and three kinds of possible degeneracies for every two bands. These three-band braiding could result in non-commutative behavior, belonging to a non-Abelian (NA) quaternion group instead of the Abelian one. Multiple two-dimensional (2D), three-dimensional (3D), and synthetic dimensional topological semimetals or complex eigenvalue space in non-Hermitian systems are commonly considered to create sufficient dimensions for band braiding, offering a convenient platform for realizing non-Abelian topological charges (NATCs) [21–31]. Conversely, few studies have explored NA band topology in one-dimensional (1D) systems due to the constrained degree of freedom in this dimension, which necessitates additional parameter space requirements, such as imaginary couplings [32].

However, 1D NA systems have their own advantages in at least two aspects compared to semimetals. One stems from their 1D BZ with periodic boundary conditions, that can be seen as a subset $k$-loop surrounding some nodal points in a parent 2D space. Consequently, the 1D NA cases naturally upgrade to global non-Abelian topological phases (NATPs), that can host endpoint states in the full band gaps, leading to more straightforward experimental observation. The other comes from their simple model, compact size, and hassle-free manipulation for future applications. A recent pioneering research demonstrated 1D NA electric circuit systems, which take advantage of its accessible complicated configurations by breaking both time-reversal and parity symmetries while keeping combined parity-time-reversal symmetry [32]. But for magnetic-free classical waves, particularly airborne sound, achieving NA behavior remains challenging.

In this paper, we report on the experimental implementation of NATPs in a 1D acoustic system. To achieve this, we incorporate an additional partner unit cell with braiding connecting tubes, which enables our acoustic system to realize positive, negative, and complex couplings within a single model. As a result, we observe all $i$, $j$, and $k$ phases, featuring endpoint states and the generation of a phase diagram. Furthermore, we verify quotient relations between different NATPs. This work extends our understanding of NA physics in classical-wave topological systems, which are well-established and mature in the design of devices.

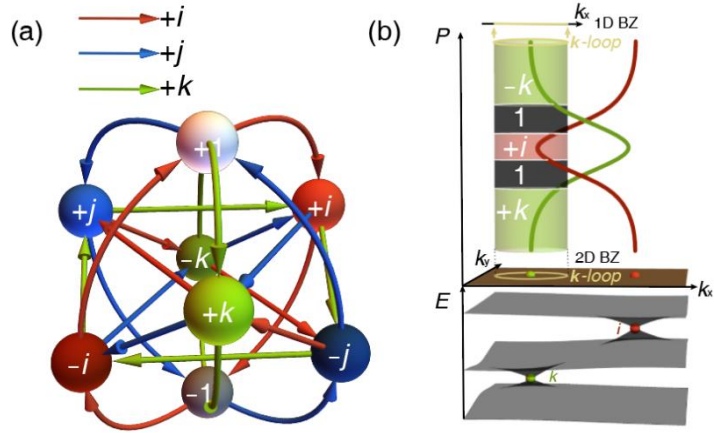

Fig. 1 | NATCs in bands. (a) Quaternion group. Balls represent group elements, and conjugate elements are in a similar color. For example, $\pm i$ are in light and dark red, respectively, while they are in the same class. Arrows in red, blue, and green define mutual multiplications by $+i/+j/+k$, respectively. For example, $(+i) \cdot$ (blue →) = $(+k)$. (b) Schematic of the NATCs. DPs $i$ and $k$ project onto the 2D BZ on a brown plane, marked as red and green dots, respectively. A $\boldsymbol{k}$-loop circling DP-$k$ in the origin changes its NATCs with $P$ increasing. A 1D BZ can be taken as such a loop in a high-dimensional space.

The NA quaternion group $\mathbb{Q} = \{+1, \pm i, \pm j, \pm k, -1\}$, shown in Fig. 1(a), is composed of three anti-commuting imaginary units that satisfy $ij = -ji = k, jk = -kj = i, ki = -ik = j$ and $i^2 = j^2 = k^2 = -1$, reminiscent of Pauli matrices. Here, NA comes from the invalid of the commutative law of multiplication $ij \neq ji$. Physically, $+1$ represents the trivial topology while $\{\pm i\}$, $\{\pm j\}$, $\{\pm k\}$, and $\{-1\}$ represent different NATCs. A physical interpretation of this group can be visualized in Fig. 1b, where two nodal points in a 2D three-band system represent degeneracies between the 2$^{nd}$ & 3$^{rd}$ and 1$^{st}$ & 2$^{nd}$ bands, respectively. Due to the symmetry, the 2$^{nd}$ & 3$^{rd}$ (1$^{st}$ & 2$^{nd}$) wave vectors acquire an additional $\pi$ phase along the $\boldsymbol{k}$-loop surrounding the nodal point, leading to the assignment of charges $i$ ($k$). As the system evolves with some parameter $P$, nodal points move along a braiding trajectory and the $\boldsymbol{k}$-loop extends to a cylinder in the $P$-BZ space. Inside the $\boldsymbol{k}$-cylinder, nodal point on every intersecting surface indicates that NATC undergoes a process of $+k \sim 1 \sim +i \sim 1 \sim -k$. The sign of $k$ changes due to the NA braiding $iki^{-1} = k^{-1} = -k$ [21], leading to the bouncing of nodal points [26–28]. The only difference between the 1D and 2D cases is that 1D BZ itself is a $\boldsymbol{k}$-loop and can be taken as a subspace in a 2D parent system. In this sense, NATC upgrades to a global NATP (yellow lines and arrow in Fig. 1b). The intersection between the nodal point trajectory and the $\boldsymbol{k}$-cylinder indicates the phase transition points, where the bandgaps close (Fig. 1b). As a result, endpoint or interface state must exist at the boundary between different NATPs.

The realization of NATC theoretically requires different types of couplings, such as positive, negative, and imaginary values [32]. It has been established that positive/negative couplings can be achieved by connecting endpoints of cavities in opposite/same directions, respectively, by utilizing acoustic dipole modes because the in-phase eigenvalue is larger/smaller than the out-of-phase eigenvalue [33]. Very recently, it has been confirmed that positive and negative couplings introduce the twisted $\pi$-flux block and equivalently complex hopping [9] (see Fig. S1 in the Supplementary Information). For practical implementation, we use a braiding formation of coupling tubes to ensure uniform length among all the tubes, and the coupling strength is controlled by their diameters. Additionally, we reshape the cavities into a U-shape, which enables tuning of on-site energy by their length and minimizes the difference between the acoustic model and the tight-binding approach (TBA) [35].

One unitcell with a lattice constant $a = 4cm$ contains 6 cavities. There are 3 cavities in one layer for the requirement of 3 bands, and a set of partners on the other layer for the construction of conjugate imaginary couplings, as shown in Fig. 2(a), with sublattice index $X_\alpha$ ($X = A, B, C$), $\alpha = \uparrow \downarrow$, respectively. The corresponding TBA model is shown in Fig. 2(b). Their on-site energies are marked as $S_X$, which is controlled by the length of the cavity $h_X$. There are no inner cell couplings between these 6 cavities. The inter cell couplings can be expressed as $v_{X_\alpha,Y_\beta} = \pm(1 - |\delta(X,Y) - \delta(\alpha,\beta)|)$, indicating that within the same layer $\alpha$, there are only couplings connecting the cavities with the same letter index $X$ ($v_X$), while between different layers, there are only couplings connecting different letter index. We further constrain that $v_{(X_\downarrow,n),(Y_\uparrow,n+1)} = -v_{(X_\uparrow,n),(Y_\downarrow,n+1)} = v_{(Y_\downarrow,n),(X_\uparrow,n+1)} = -v_{(Y_\uparrow,n),(X_\downarrow,n+1)} \equiv v_{XY}$, with $n$ marks the unit cell. All connecting tubes have the same length, and the strength of coupling is determined by their diameter $d_X/d_{XY}$. In practice, we use the sign "$\pm d$" to indicate positive/negative couplings and mark them as red/blue in Fig. 2(a). The TBA Hamiltonian in $k_x$ space can be written as

$$H(k_x) = \begin{bmatrix} H_d(k_x) & iH_o(k_x) \\ -iH_o(k_x) & H_d(k_x) \end{bmatrix}$$

$$H_d(k_x) = \begin{bmatrix} S_A + 2v_A \cos k_x & 0 & 0 \\ 0 & S_B + 2v_B \cos k_x & 0 \\ 0 & 0 & S_C + 2v_C \cos k_x \end{bmatrix} \quad (1)$$

$$H_o(k_x) = 2\sin k_x \begin{bmatrix} 0 & v_{AB} & v_{CA} \\ v_{AB} & 0 & v_{BC} \\ v_{CA} & v_{BC} & 0 \end{bmatrix}$$

with the basis $(A_\uparrow, B_\uparrow, C_\uparrow, A_\downarrow, B_\downarrow, C_\downarrow)$. With the unitary transformation $U = \exp(i\tau_x \otimes \sigma_0 \pi/4)$, in which $\tau_x$ is the x-component of Pauli matrix and $\sigma_0$ is the $3\times 3$ identical matrix, the Hamiltonian can be diagonalized with new basis $(A_\uparrow + iA_\downarrow, B_\uparrow + iB_\downarrow, C_\uparrow + iC_\downarrow, A_\uparrow - iA_\downarrow, B_\uparrow - iB_\downarrow, C_\uparrow - iC_\downarrow)/\sqrt{2}$:

$$H_0(k_x) = UH(k_x)U^\dagger = \begin{bmatrix} H_d(k_x) + H_o(k_x) & 0 \\ 0 & H_d(k_x) - H_o(k_x) \end{bmatrix} \equiv \begin{bmatrix} H_+(k_x) & 0 \\ 0 & H_-(k_x) \end{bmatrix} \quad (2)$$

$H_\pm(k_x)$ are Hamiltonian with a pair of conjugate NATCs [32]. The effective model is shown in Fig. 2(c).

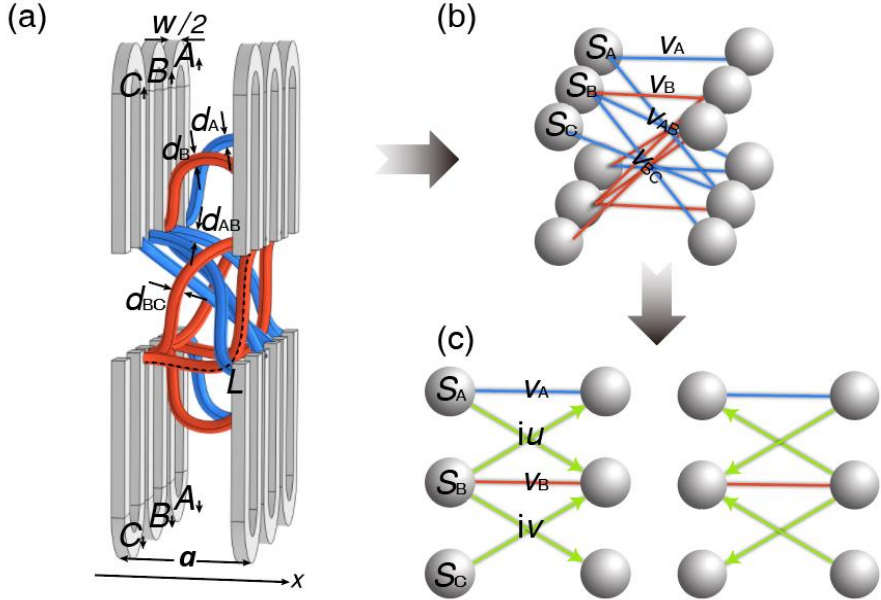

Fig. 2. Model construction. (a) Two layers of acoustic structure. U-shaped cavities are connected with effective positive couplings by red tubes, and negative couplings by blue tubes. All the tubes share the same length $L = 7cm$. (b) TBA model with pure real couplings. (c) Effective TBA model with imaginary couplings.

We can calculate the NA charge by:

$$Q = \overline{\exp}\left[\oint A(k_x) \cdot dk_x\right] \tag{3}$$

where $A(k) = \sum_t \beta_t \left(-\frac{i}{2}\sigma_t\right)$ with $\sigma_t$ as the Pauli Matrix, $\beta_t = \sum_{pq}\langle u^p|\partial_{k_x}|u^q\rangle\epsilon_{tpq}/2$ calculated from bands $p$ and $q$, $\epsilon_{tpq}$ being fully antisymmetric tensor. The bar over the exponent indicates path ordering. The final results of quaternion group elements are represented as Pauli matrix $1 \to \sigma_0$, $i \to i\sigma_x$, $j \to i\sigma_y$, and $k \to i\sigma_z$, where $\sigma_0$ is the 2 by 2 identify matrix. It should be noticed that the signs $\pm$ only have relative meaning for NATCs $i$, $j$, and $k$, but distinguish the trivial phase 1 and nontrivial phase $-1$ for the unit 1. $i$, $j$, and $k$ can only be distinguished up to a basepoint, which means that a pair of conjugate charges cannot be distinguished by observing them separately. Their difference only manifests as interface states sandwiched by them [32]. In this sense, although our design cannot distinguish a pair of conjugate NATCs, it provides a good platform to study one single phase and the interface between phases in different classes. In the following, we use the NATC of $H_+(k_x)$ to represent a specific system.

The 9 parameters in $H_+$ can be classified into 3 groups: 3 on-site energies $S_{A,B,C}$, 3 intercell couplings with the same letter index $v_X$, and 3 intercell couplings with different letter index $v_{XY}$. The first two groups are in the diagonal position of the Hamiltonian. The sum of each group ($S_A + S_B + S_C$ & $v_A + v_B + v_C$) does not affect the NATC, because they can only shift or scale the bands. To ensure orthogonality, we set the other parameters as $S_A - S_B \equiv S_-$, $S_A + S_B - 2S_C \equiv S_+$, $v_A - v_B \equiv v_-$, $v_A + v_B - 2v_C \equiv v_+$, $v_{AB} \equiv u$, $v_{BC} \equiv v$, and $v_{CA} \equiv w$. Without the loss of generality, we neglect the next nearest coupling ($w = 0$) and the difference of the first two sites ($S_- = 0$), and set $u = 1$ as the unit. We obtain a phase diagram with 3 of remaining parameters: $v_-$, $v_+$ and $S_+$ as shown in Fig. 3(a) leaving $v = 1$. The bandgap closes at phase transition surfaces.

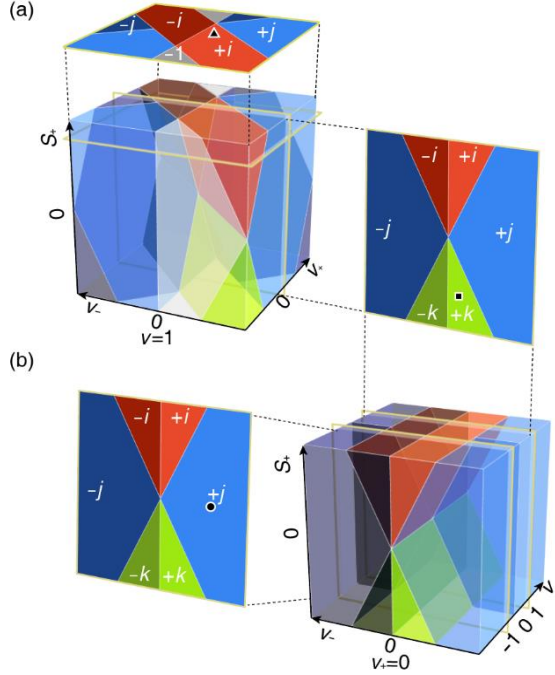

Fig. 3. NATP diagram. (a) NATP diagram spanned by parameters $v_\pm$ and $S_+$. $\pm i$, $\pm j$, $\pm k$ and $-1$ phases are in red, blue, green and grey, respectively. Slices of $v = 1$ and $v_+ = 0$ are picked to mark representative states $i$ and $k$ by square and rectangular, respectively. (b) Phase diagram spanned by parameters $v$, $v_-$ and $S_+$. Its $v = 1$ slice is just the $v_+ = 0$ in (d). Slice $v = -1$ is selected to mark representative state $j$ by a circle.

1D NATPs feature endpoint states in bulk gaps. We first select states in $i$ and $k$ phases marked as triangular and square, respectively, in Fig. 3(a) (Details of parameters can be found in Table I in the Supplementary Information). The sample of phase $i$ with probe and source is presented in Fig. 4(a) (See Fig. S2 in the Supplementary Information for details). In the frequency range we focus on, there are three bands with two gaps. Non-trivial NATPs can be observed by the rotation of the state when wave vector $k_x$ goes through the loop $-\pi/a$ to $\pi/a$. For instance, in the $i$ phase, the states of the 2nd and 3rd bands get $\pi$ phase while the lowest state maintains its sign (See Fig. S3 in the Supplementary Information for details). Consequently, we can see an endpoint state between the twisting bands. The experiment results are shown in Fig. 4(b), which agree with the simulation and TBA results (See Fig. S4 in the Supplementary Information for details). The states in the $k$ phase twist its 1st and 2nd bands and manifest an endpoint state between them as shown in Fig. 4(c).

The Phase diagram provides us with more information when the fourth dimension is introduced. By fixing $v_- = 0$, and treating inactive $v$ as an axis, we obtain the phase diagram in Fig. 3(b). It is worth noting that the slice of $v=1$ in Fig. 3(b) is identical to the slice $v_-=0$ in Fig. 3(a). The phases appear to be unaffected along $v$-axis, except for the band degeneracy at $v=0$, which does not lead to a phase transition. We select a state in the $j$ phase denoted by a circle, which twist its 1st and 3rd bands, and two set of states sandwich the 2nd band can be observed. It should be noted that these in-gap states are not guaranteed by the NA topology, but rather emerge due to specific parameter choices. They could be even tuned into the 2nd bulk band, forming bound-states-in-continuum [32].

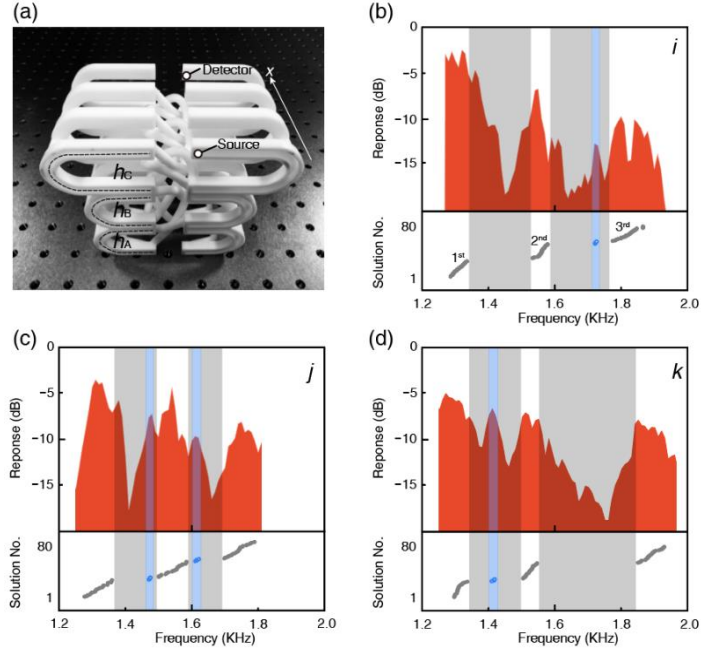

Fig. 4. Endpoint states of a single phase. (a) A sample of phase *i* and the experimental setup to detect the response. Parameters are shown in Table. I in the Supplementary Information. (b-d) Responses of phases *i, j* and *k*, respectively. Simulated solutions are shown in the lower panel.

In the NA case, rather than bulk-boundary correspondence, NA quotient relations can be observed at the domain wall of two different NATPs $Q_L$ and $Q_R$ with the topology $\Delta Q = Q_L Q_R^{-1} = Q_L/Q_R$ [32]. Specifically, the domain-wall between phases *j*&*k* (*k*&*i*/*i*&*j*) acts similarly the endpoint in one single phase *i* (*j*/*k*). In fact, in Abelian cases, e.g. Chern insulator, the interface state can be determined by the difference of Chern numbers on its two sides $\Delta C = C_L - C_R$, which can be taken as the Abelian group multiplication product of $C_L$ and $C_R^{-1}$, and $\Delta C = C_L C_R^{-1}$. Therefore, both Abelian and non-Abelian cases follow the same rule that the domain-wall state is determined by the original and inverse value of NATPs on its two sides, respectively. The experimental sample with the detector and source marked is shown in Fig. 5(a). NA quotient relations are visible in our acoustic systems, as shown in Figs. 5(b-d).

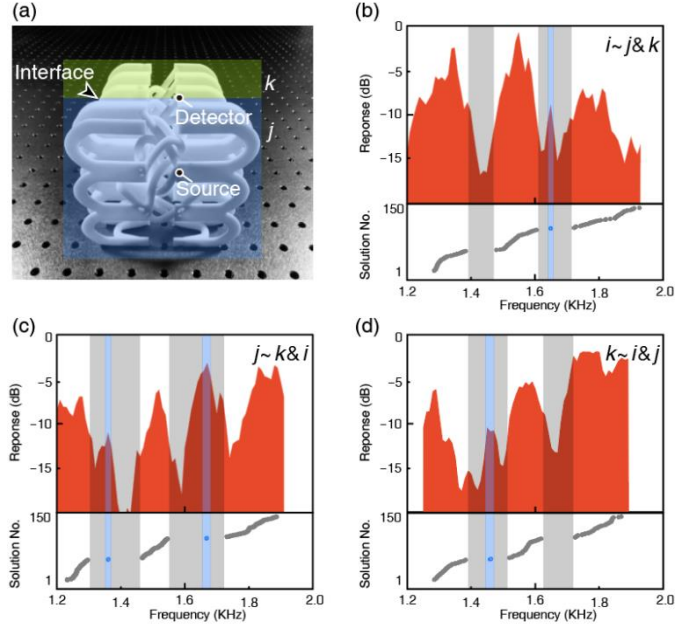

Fig. 5. Interface states between different phases. (a) A sample of phase *i vs k* and the experimental setup to detect the response. Parameters are shown in Table. I in the Supplementary Information. (b-d) Responses of interfaces *j vs k*, *k vs i* and *i vs j*, respectively. Simulated solutions are shown in the lower panel.

In summary, our work demonstrates the realization of non-Abelian topological phases in acoustic systems, opening up new avenues for exploring bosonic non-Abelian topological phases. This can be further combined with non-Hermitian [24] and Floquet topology [36] concepts, or other non-Abelian braiding process [37–41] to explore new topological physics. In contrast to its 2D and 3D counterparts, 1D NA topology exhibits a distinct endpoint state in a complete gap, making it a promising candidate for designing functional primitive structures. By utilizing acoustic dipole modes, we effectively implement positive, negative, and equivalently complex couplings, making acoustic crystals a versatile platform for simulating almost all kinds of TBA models in theory. Our design only requires one auxiliary unit cell and provides probably the simplest method to achieve effective complex couplings in practice, which could be useful for compact devices. The connecting tubes can be replaced by soft tubes, making it a plug-and-play and tunable topological device for future quantum computing designs. The phase diagrams obtained with different parameters reveal the system's distinct physical properties, directly indicating the topological intricacy.

**Acknowledgements**

The work was jointly supported by the National Key R&D Programme of China [Grant No. 2022YFA1404302 (C.H.)] and the National Natural Science Foundation of China [Grant Nos. 52022038 (C.H.), 92263207 (C.H.), 11874196 (C.H.), 11890700 (Y.-F.C.), 51721001 (Y.-F.C.), 52027803 (Y.-F.C.), and 52103341 (X.-C.S.)].

**Author contributions**

X.-C. S. and C. H. conceived the original idea. X.-C. S. performed the theoretical portions of this work. J.-B. W. conducted the experiments. C. H. and Y.-F. C. supervised the project. All authors contributed to the analyses and the preparation of the paper.

**Competing interests**

The authors declare no competing interests.

Supplementary Information for **"Non-Abelian Topological Phases and Their Quotient Relations in Acoustic Systems "**

Xiao-Chen Sun[1,2,§], Jia-Bao Wang[1,§], Cheng He[1,2,3,*], Yan-Feng Chen,[1,2,3,†]

[1]National Laboratory of Solid State Microstructures & Department of Materials Science and Engineering, Nanjing University, Nanjing 210093, China

[2]Collaborative Innovation Center of Advanced Microstructures, Nanjing University, Nanjing 210093, China

[3]Jiangsu Key Laboratory of Artificial Functional Materials, Nanjing University, Nanjing 210093, China

*Corresponding author.

chenghe@nju.edu.cn

†Corresponding author.

yfchen@nju.edu.cn

§ X.-C. S., J.-B. W. contributed equally to this work.

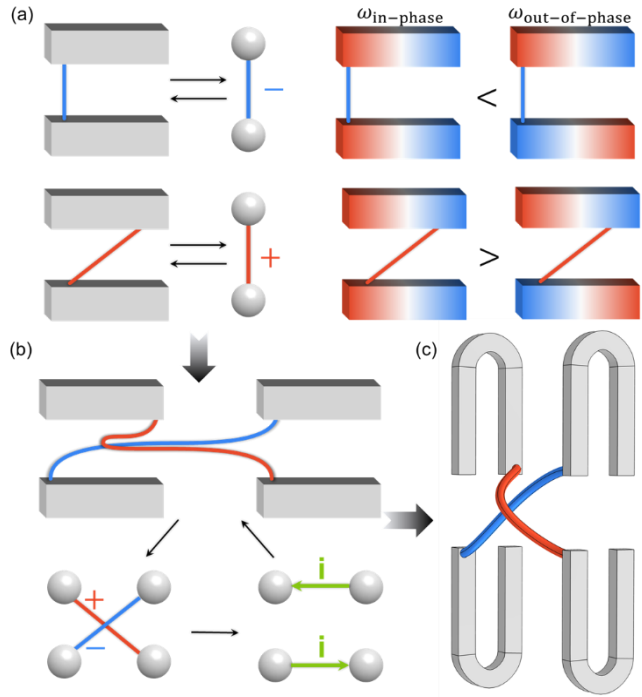

Fig. S1 | Construction of various effective couplings with acoustic dipole modes. (a) Effective positive and negative couplings. When connected with tubes, the degeneracy of the dipole modes split into in-phase and out-of-phase modes. Effective positive coupling leads to the larger in-phase mode eigen frequency, while negative coupling leads to larger out-of-phase frequency. They can be realized by connecting endpoints of cavities of opposite and same direction, respectively. (b) Effective imaginary coupling. Cavities in the double-layer can be connected by crossing positive and negative couplings. With the change of basis, it realizes effective complex coupling (pure imaginary coupling in this case). (c) U-shaped cavities connected by braiding tubes with the same length in experiments.
16

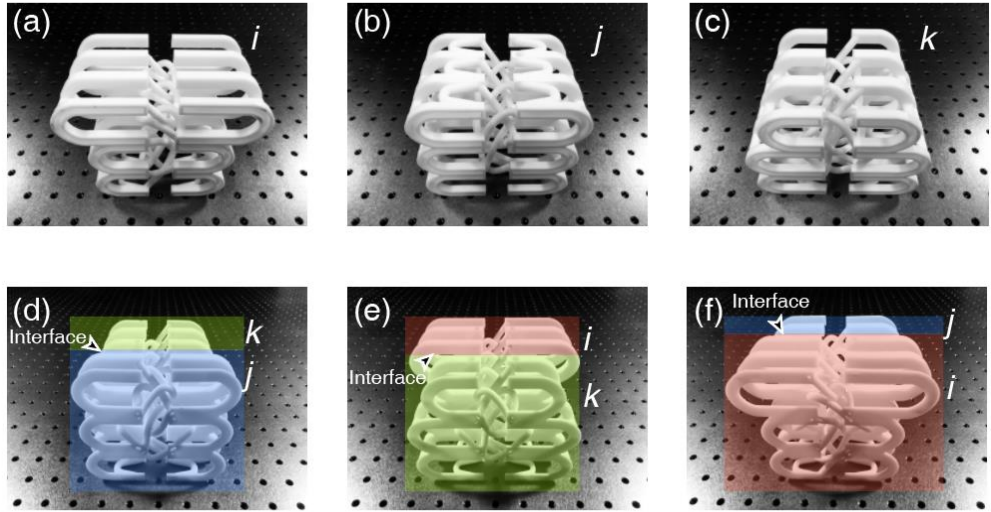

Fig. S2 | Samples in experiments. (a-c) Experimental samples for endpoint responses for the *i*, *j* and *k* phases. (d-f) Experimental samples for interface responses for the *j&k*, *k&i* and *i&j* phases. *i*, *j* and *k* phases are marked as red, blue, and green, respectively.

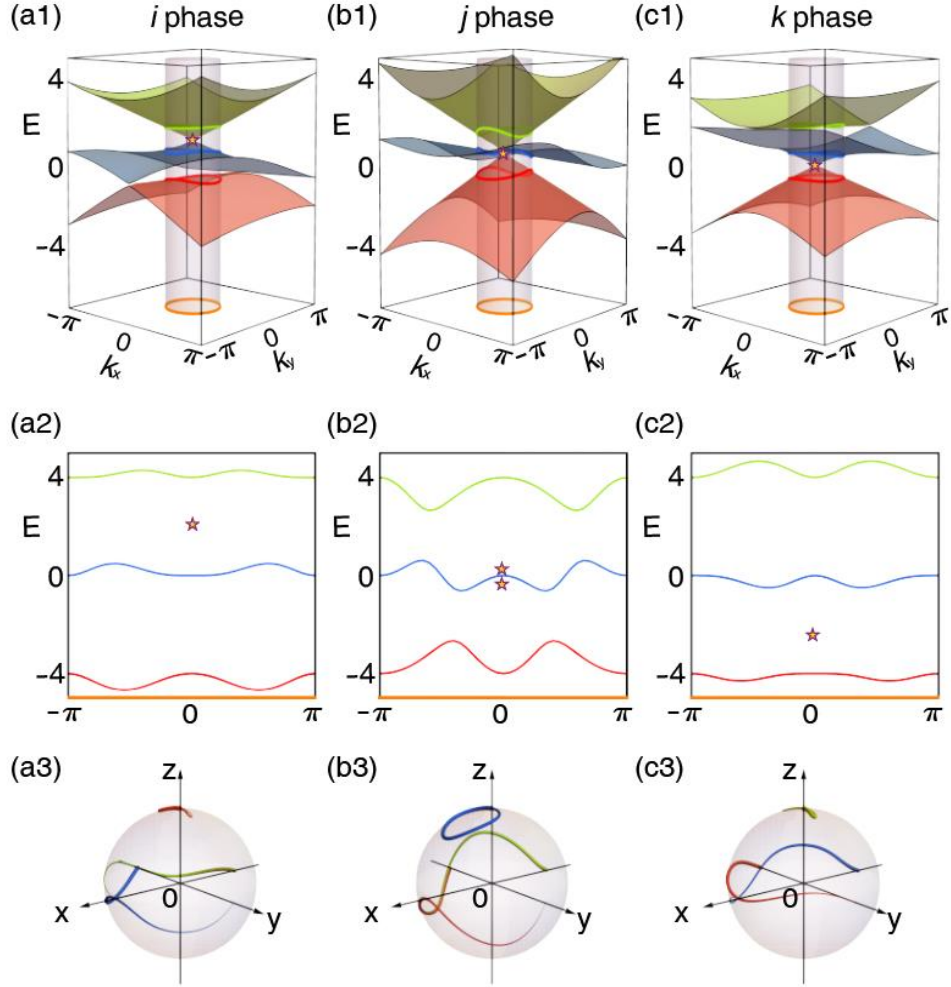

Fig. S3 | Eigenstates rotation. (a1-c1) 2D band structure, with 1$^{st}$ 2$^{nd}$ and 3$^{rd}$ bands marked as red, blue and green, respectively. The Hamiltonian is Eq. (1) in the manuscript with the replacements: $\cos k \to k_x, \sin k \to k_y$. Stars mark degenerate points. (a2-c2) 1D band structure. It can be taken as a subspace as the orange circle shown in Fig. (a1-c1). Stars mark the endpoint states matching degenerate states in Fig. (a1-c1). (a3-c3) The rotation of eigenstates in 1D system. The thickness gradient marks the eigenstates varying with $k$ from $-\pi$ to $\pi$ (taken $a = 1$). It can be seen that eigenstates of 1$^{st}$(red), 2$^{nd}$(blue) and 3$^{rd}$(green) are almost static in $i$, $j$, and $k$ phases, respectively, while the other two states rotate around it by $\pi$ phase.

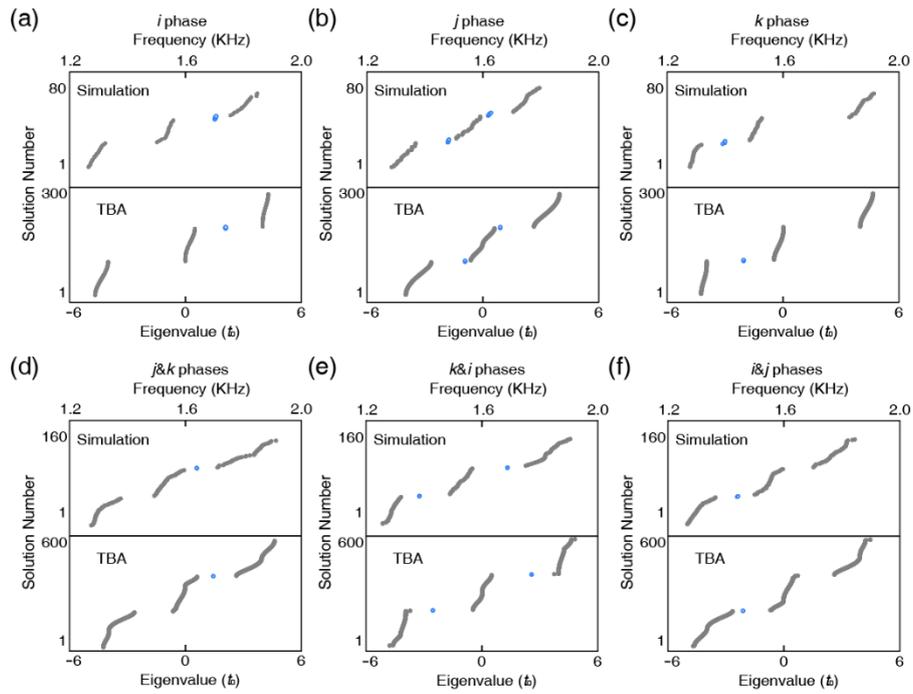

Fig. S4 | Endpoint and interface states from simulation and TBA, respectively.

TABLE I. Parameters for various phases

| Phases | Methods | $S_A/h_A$ | $S_B/h_B$ | $S_C/h_C$ | $v_A/d_A$ | $v_B/d_B$ | $u/d_{AB}$ | $v/d_{BC}$ |
|---|---|---|---|---|---|---|---|---|
| $i$ | TBA | 2 | 2 | −4 | −1 | 1 | 1 | 1 |
| | Simulation (cm) | 4.84 | 4.84 | 6.68 | −0.48 | 0.48 | 0.48 | 0.48 |
| $j$ | TBA | 0 | 0 | 0 | −2 | 2 | 1 | −1 |
| | Simulation (cm) | 5.5 | 5.5 | 5.5 | −0.72 | 0.72 | 0.48 | −0.48 |
| $k$ | TBA | −2 | −2 | 4 | −1 | 1 | 1 | 1 |
| | Simulation (cm) | 6.23 | 6.23 | 4.39 | −0.48 | 0.48 | 0.48 | 0.48 |

Note: "−" in Simulation means that the effective negative coupling is used.

## Experimental measurement and numerical simulation

All samples used in the experiments are fabricated using photosensitive resin (Godart$^{TM}$8111X) via 3D printing (geometry tolerance of 0.1 mm). This stereolithography material (modulus 3160 MPa, density $1.14 \text{gcm}^{-3}$) is regarded as an acoustic hard boundary for the impedance mismatch. The thicknesses of all sites and tube walls are 1 mm. In the experiments, we drill some round holes on the cavities at the upper and lower surfaces and print a corresponding plug for the convenience of measurements. In the measurements, commercial loudspeakers (AMT-47) and microphones (BSWA MPA416) are used as the acoustic source and detector, respectively.

Full-wave simulations are implemented by the commercial software COMSOL Multiphysics with a 3D acoustic module based on a finite element method. The mass density and sound velocity are $1.21 \text{kgm}^{-3}$ and $343 \text{ms}^{-1}$, respectively.